\documentclass[twocolumn,
prl
,superscriptaddress
]{revtex4}

\usepackage{graphicx}
\usepackage{amssymb}
\usepackage{amsmath}
\usepackage{color}
\usepackage{comment}
\usepackage[normalem]{ulem}

\usepackage{tikz}
\usetikzlibrary{decorations.markings,arrows}
\tikzset{
    fermi/.style={draw=black, postaction={decorate},
        decoration={markings,mark=at position .55 with {\arrow[draw=black]{>}}}}
     }

\newcommand{\vect}[1]{{\mathbf #1}}
\newcommand{\ad}{a_{2\rm{D}}}

\newcommand{\up}{\uparrow}
\newcommand{\down}{\downarrow}
\renewcommand{\k}{{\bf k}}
\newcommand{\eb}{\varepsilon_B}
\newcommand{\p}{{\bf p}}

\newcommand{\ef}{\varepsilon_F}

\newcommand{\ek}{\epsilon_{\k}}

\newcommand{\nn}{\nonumber}

\def \del{\partial}    

\begin{document}


\title{Pair Correlations in the Two-Dimensional Fermi Gas}

\author{Vudtiwat Ngampruetikorn}
\affiliation{T.C.M. Group, Cavendish Laboratory, JJ Thomson Avenue, Cambridge,
 CB3 0HE, United Kingdom} %

\author{Jesper Levinsen}
\affiliation{T.C.M. Group, Cavendish Laboratory, JJ Thomson Avenue, Cambridge,
 CB3 0HE, United Kingdom} %
\affiliation{Aarhus Institute of Advanced Studies, Aarhus University, DK-8000 Aarhus C, Denmark}

\author{Meera M. Parish}
\affiliation{London Centre for Nanotechnology, Gordon Street, London, WC1H 0AH, United Kingdom}

\date{\today}

\begin{abstract}
We consider the two-dimensional Fermi gas at finite temperature with attractive short-range interactions. 
Using the virial expansion, which provides a controlled approach at
high temperatures, we determine the spectral function and 
contact for the normal state.  Our calculated spectra are in
qualitative agreement with recent photoemission measurements~[M. Feld
et al., Nature \textbf{480}, 75 (2011)], 
thus suggesting that the observed pairing gap is a feature of the
high-temperature gas rather than being evidence of a pseudogap regime
just above the superfluid transition temperature. We further argue
that the strong pair correlations result from the fact that the
crossover to bosonic dimers occurs at weaker interactions than
previously assumed.
\end{abstract}


\maketitle

The two-dimensional (2D) Fermi gas with contact interactions provides a basic model for understanding pairing and superconductivity in 2D systems~\cite{Randeria1989,Randeria1990,SR1989}. Recently, there has been a resurgence of interest in this model owing to the
realization 
of 2D Fermi gases in cold-atom experiments~\cite{Martiyanov2010,Feld2011,Frohlich2011,Dyke2011,Sommer2012,Zhang2012,Baur2012,Koschorreck2012,Frohlich2012,Makhalov2013}. 
In particular, there is the possibility of investigating deviations from Fermi liquid behavior 
at finite temperature. 
However, 2D systems are notoriously difficult to treat theoretically since mean-field theory is less reliable in low dimensions, while at the same time there is a dearth of exact solutions, in contrast to the case in 1D. Hence there is a need for controlled perturbative approaches 
that can guide theory and experiment.
The virial expansion is one such controlled approach in the high temperature limit, and this will be 
the focus of our Letter.

Before investigating finite temperature, it is instructive to consider the ground state of the two-component  ($\up,\down$) Fermi gas in 2D. Here we assume that the masses and chemical potentials of each spin are equal, i.e.,  $m_\up=m_\down \equiv m$ and $\mu_\up=\mu_\down \equiv \mu$, respectively. 
Then one expects a smooth crossover from BCS pairing to Bose-Einstein condensation (BEC) of tightly bound $\up\down$ dimers with increasing interspecies attraction. 
Since there is always a two-body bound state for attractive contact interactions in 2D, one can define the strength of the interaction using the two-body binding energy $\eb = \hbar^2/ma_\text{2D}^2$, where $a_\text{2D}$ is the 2D scattering length. Thus, for a Fermi gas with total density $n = k_F^2/2\pi$, where $k_F$ is the Fermi wave vector, we have $k_F\ad \gg 1$ in the BCS limit and
$k_F\ad \ll 1$ in the BEC limit.
Note that the crossover can be driven by varying the density as well as by varying the interaction, unlike the case in 3D.

Mean-field theory has provided a qualitative understanding of the 2D BCS-BEC crossover~\cite{Randeria1989,Randeria1990}, while recent quantum Monte Carlo (QMC) calculations have determined the ground state energy~\cite{Bertaina2011}.
However, there is surprisingly little discussion about what constitutes the \emph{crossover point}.
As previously argued in the 3D case, the crossover is best defined as the point where $\mu=0$, since this can 
signal the disappearance of a Fermi surface and an associated change in the quasiparticle excitation spectrum~\cite{Leggett1980,Parish2005}.
Indeed, while the $s$-wave paired superfluid is characterized by a crossover,
a phase transition at $\mu=0$ is predicted in the $p$-wave paired superfluid~\cite{Gurarie2007}.
In 2D, it is generally assumed that the crossover occurs at $\ln(k_F\ad) = 0$, which is consistent with the mean-field chemical potential, $\mu = \ef -\eb/2$, being zero (where $\ef = \hbar^2 k_F^2/2m$ is the Fermi energy). However, it is easy to show using the QMC data from Ref.~\cite{Bertaina2011} that 
the chemical potential in fact vanishes 
at $\ln(k_F\ad) \simeq 0.5$, corresponding to $n \ad^2 \simeq 0.4$.
This suggests that experiments on the  2D Fermi gas have mainly 
explored the BEC side of the crossover, as we discuss below. 
Therefore, it is perhaps not surprising that recent radio-frequency (rf) measurements in the regime  $\ln(k_F\ad) \lesssim 0.6$ agree well with two-body theory~\cite{Sommer2012}.

This also has implications for the normal state above the superfluid
transition temperature $T_c$.  In particular, the highly sought-after
``pseudogap'' regime requires the presence of a Fermi
surface~\cite{Trivedi1995}: while this phase is often synonymous with
pairing above $T_c$ in the cold-atom literature, if one is to emulate
the pseudogap phenomena in high temperature superconductors, then one
must have a loss of spectral weight, i.e., a gap, at the Fermi
surface~\cite{Loktev2001}.  Thus, a pairing gap above $T_c$ does not
imply a pseudogap when $\mu <0$ and Pauli blocking is minimal.

In this Letter, we investigate the 2D Fermi gas at finite temperature
$T$ using the virial expansion. We focus on the pair correlations as
encoded in the contact~\cite{Tan2008} and the spectral function,
rather than the thermodynamic properties and the equation of state,
which have been determined in Ref.~\cite{Liu2010}.  We find that our
calculated spectra qualitatively reproduce those from recent
photoemission experiments~\cite{Feld2011}, and we argue that the
observed pairing gap above $T_c$ is in the regime where $\mu<0$.

The basic idea of the virial expansion, as applied to 
the uniform 2D Fermi gas, can be summarised as follows. 
Working in the grand canonical ensemble, we define the
virial coefficients $b_j$ such that the grand potential
$\Omega(T,V,\mu)$ is given by (we set $\hbar=k_B=1$)
\begin{equation}
\Omega = -2TV\lambda^{-2} \sum_{j\ge1} b_jz^j,
\label{eq:Pvirial1}
\end{equation}
where the thermal wavelength $\lambda = \sqrt{2\pi/mT}$ and the
fugacity $z=e^{\beta \mu}$. We also denote $V$ as the system area and
$\beta \equiv 1/T$.  In the high temperature limit where
$\lambda^2n\lesssim 1$, the above power series may be truncated and
the thermodynamics of the system can be accurately described by only
the first few virial coefficients.  This usually corresponds to an
expansion in small $z$, but care must be exercised when treating the
Bose limit $k_F\ad \ll 1$. In this case, $\mu \simeq -\eb/2$ at low
temperatures so that $z \simeq e^{-\beta \eb/2} \to 0$ as $T \to 0$,
which naively suggests that the virial expansion is valid even at zero
temperature.  However, as we show below, the coefficients $b_j$ also
contain powers of $e^{\beta \eb/2}$ that cancel the contribution from
the binding energy in $z$ when $j$ is even. Thus, the relevant
expansion parameter is in fact $z^\text{(Bose)} =z e^{\beta \eb/2}$,
with corresponding coefficients $b_j^\text{(Bose)} = e^{-j\beta \eb/2}
b_j$.

\begin{figure}
\centering
\includegraphics[width=.9\linewidth]{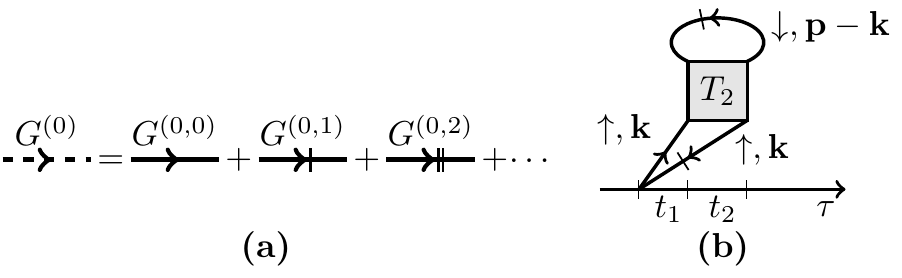}
\caption{(a) Diagramatic representation of the virial expansion of the
  bare propagator. The number of vertical dashes indicates the power
  of fugacity of the corresponding term in
  Eq.~(\ref{eq:virialexp}). (b) Second order contribution to
  density due to interactions. $t_1$ and $t_2$ represent
  intervals on the imaginary time axis. 
  \label{fig:Gvirial}}
\end{figure}

To proceed further, we consider the total density
\begin{equation}
n = -\frac{1}{V} \left.\frac{\partial \Omega}{\partial \mu}\right|_{T,V} 
= 2\lambda^{-2} \sum_{j\ge1} jb_jz^j,
\label{eq:dens}
\end{equation}
with the density of each species given by $n_\sigma = n/2$.  The
non-interacting part of the virial coefficient is then readily
obtained from $n_\sigma =\sum_\k n_F(\ek-\mu)$, where $n_F$ is the
Fermi-Dirac distribution function. This yields
%
$b_j^\text{(free)} = (-1)^{j-1}j^{-2}$ 
%
for $j\ge1$.
We obtain the second and third virial coefficients using the
diagrammatic method recently introduced in 3D by
Leyronas~\cite{Leyronas2011}.  As a starting point, we write down the
relation between the density and the propagator,
%
$n_\sigma(\mu,T) =  \sum_{\k} G_\sigma(\k,\tau=0^-)$.
%
As usual, the full Green's function on the right-hand side  is then
written in terms of the bare propagators $G^{(0)}(\k,\tau)$. The key
point is that these may be expanded in powers of $z$,
\begin{align}
G_\sigma^{(0)}(\k,\tau)
& = e^{-(\epsilon_{\k}-\mu)\tau}[-\Theta(\tau)+n_F(\epsilon_{\k}-\mu)]\\
&= e^{\mu \tau} \sum_{j\ge0} G_\sigma^{(0,j)}(\k,\tau)z^j , \label{eq:virialexp}
\end{align}
where the `virial propagators' are defined as
\begin{align}
G_\sigma^{(0,j)}(\k,\tau) =\left\{ 
  \begin{array}{l l}
    -\Theta(\tau) e^{-\epsilon_{\k}\tau}& \, \text{if} \quad j=0\\
    (-1)^{j-1} e^{-\epsilon_{\k}\tau-j\beta\epsilon_{\k}} & \, \text{if} \quad j\ge1
  \end{array} \right.
\end{align}
This allows one to construct a perturbative expansion in $z$ for the
full propagator.  As an example, Fig.~\ref{fig:Gvirial}(a) depicts the
diagrammatic representation of the virial expansion of the free
propagator.  The non-interacting virial coefficients
are also easily recovered using the virial
propagators above. Note that $G^{(0,0)}$ is the only retarded term,
meaning it vanishes whenever $\tau<0$.

\begin{figure}
\centering
\includegraphics[width=0.8\linewidth]{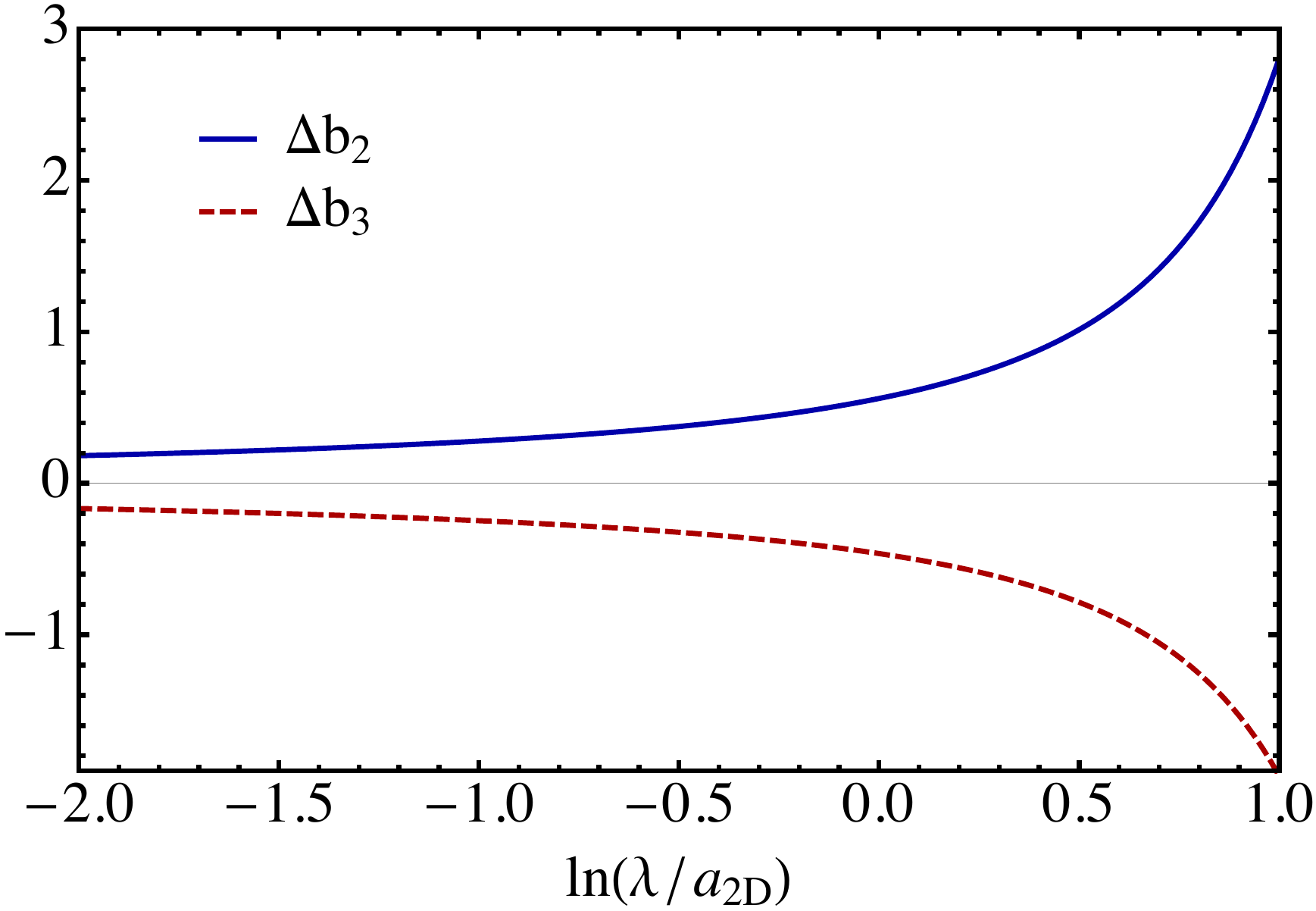}
\caption{The contribution from interactions to the second and third
  virial coefficients of the uniform 2D Fermi gas.
\label{fig:virial}}
\end{figure}

In order to obtain the second virial coefficient, we must calculate the
term arising from interactions and contributing to the density at
second order in the fugacity:
\begin{align}
2z^2\sum_{\k\p}\int & \text dt_1\text dt_2\, T_2(\p,t_2)
G_\up^{(0,0)}(\k,t_1)\nn \\ &\times 
G_\up^{(0,1)}(-\k,-t_1-t_2)G_\down^{(0,1)}(\p-\k,-t_2),
\end{align}
as illustrated in Fig.~\ref{fig:Gvirial}(b). The factor of 2 accounts
for the spin degeneracy. The $T$-matrix $T_2$ \footnote{The $T$-matrix
  in imaginary time is related to the usual frequency-dependent expression 
$\tilde
  T_2(\omega) = -\frac{4\pi}{m} \frac1{\ln(-\omega/\eb)}$ via a
  Laplace transformation $T_2(\k,\tau)=\int\frac{d\omega}{2\pi
    i}e^{-\omega\tau}
\tilde T_2\left(\omega-\frac{k^2}{2M}\right).$ }
is defined
purely in terms of the retarded virial propagator $G^{(0,0)}$ and as a
result $T_2$ is also retarded.
Following manipulations of the time integrations along the lines set
out in Ref.~\cite{Leyronas2011}, we find that the second virial
coefficient consists of two terms. The first arises from the dimer
pole of the $T$-matrix and is given by $e^{\beta\eb}$, while the
second comes from the scattering states. We then obtain
\begin{align}
\Delta b_2=e^{\beta \eb}-\int_0^\infty \frac{\text dp}{p}\, 
    \frac{2e^{-\beta p^2/m}}{\pi^2 + 4\ln^2(pa_\text{2D})},
\end{align}
where we define $\Delta b_j\equiv b_j-b_j^\text{(free)}$.  In the
limit $T\ll\eb$, the exponential term dominates as expected, since all
the atoms are then bound into pairs.
We see that, as the virial expansion contains terms
  contributing separately to the atom and dimer densities, measuring
  these in experiment may be used to extract the chemical potential and
  temperature, within the regime of validity of the approximation.
The dimer pole of the $T$-matrix
also plays a major role in the higher order coefficients $b_j$,
contributing $e^{j\beta\eb/2}$ for even $j$ and $e^{(j-1)\beta\eb/2}$
for odd $j$. Comparing this with the factors of $e^{-\beta\eb/2}$ in
$z$ discussed above, we thus expect odd terms in the expansion to tend
to zero as $\beta\eb \to \infty$. Beyond that, the calculation of the
third virial coefficient is significantly more involved, and we refer
the interested reader to Ref.~\cite{Leyronas2011} for the required
diagrams.

Figure~\ref{fig:virial} displays the behavior of the second and third
virial coefficients in the regime of interactions before
$e^{\beta\eb}$ dominates. We stress that the virial coefficients are
functions of $\ln(\lambda/\ad)$, or equivalently $\beta\eb$, only.
We see that the correction to the second virial coefficient due to
interactions is attractive, since it increases the density at fixed
$\mu$ and $T$.  However, this lowest order term is expected to
overestimate the attraction at lower temperatures and thus the
third-order correction acts to suppress the density.

We now turn to the investigation of pair correlations, for which we
first discuss the virial expansion of the contact~\cite{Tan2008} in
2D. Within the zero-range model, this quantity is given by
%
$\mathcal C = 2\pi m \frac{\text d E}{\text d \ln(\ad)}$~\cite{Werner2012}.
%
The contact can also be expressed in terms of the grand 
potential $\Omega(T,V,\mu)$, thus giving 
\begin{align}
\mathcal C = 2\pi m
   \left(\frac{\partial \Omega}{\partial
       \ln(\ad)}\right)_{T,V,\mu} 
=\frac{8\pi^2V}{\lambda^4}
\sum_{j\ge2} c_jz^j,
\end{align}
with `contact' coefficients $c_j =[\text d \Delta b_j/\text d
\ln(\lambda/\ad)]_T$ (for the 3D equivalent, see
Ref.~\cite{LiuReview}).

\begin{figure}
\centering
\includegraphics[width=0.8\linewidth]{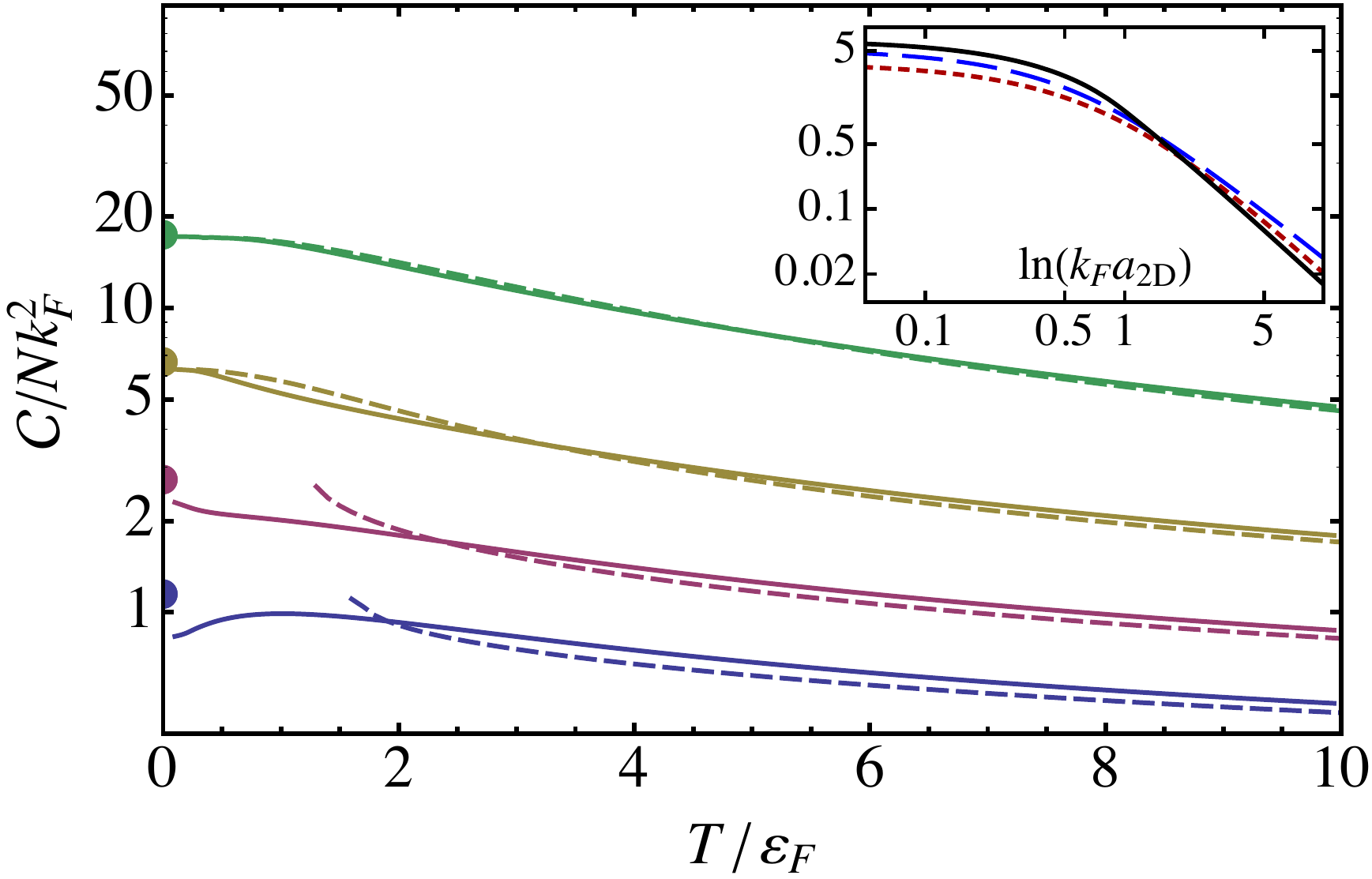}
\caption{The contact calculated within the second (solid) and third
  (dashed) order of the virial expansion for
  $\ln(k_F\ad)=-0.5,0,0.5,1$ (top to bottom).  The filled circles
  are the results of zero-temperature
  QMC~\cite{Bertaina2011}. Inset: Contact from $T=0$ QMC
    (solid), the virial expansion up to second order at $T=\ef$
    (dashed), and at $T=3\ef$ (dotted).
\label{fig:contact}}
\end{figure}

In Fig.~\ref{fig:contact} we display the contact as a function of
temperature $T/\ef$ for different $\ln(k_F\ad)$.  Over a large
temperature range, we observe a very good agreement between the second
and the third order of the virial expansion. Moreover, the results
from QMC at $T=0$~\cite{Bertaina2011} match our calculated contact
reasonably well.  However, the expansion starts to break down at low
temperatures once we move toward the BCS limit.  Interestingly, it
appears that the second order of the expansion extrapolates better
between the high and low temperature limits than the higher order
expansion.  
As shown in the inset of Fig.~\ref{fig:contact}, the contact
  is non-monotonic with temperature when $\ln(k_F\ad) \gtrsim 1$. This
  suggests a way of distinguishing the BCS regime from the Bose limit
  at intermediate temperatures.
  {We speculate that the contact initially increases with temperature in the BCS regime as Pauli blocking becomes less important.}
  For a discussion of non-monotonicity of the 3D contact, see Ref.~\cite{Yu2009}.
In the limit $T \gg \eb$, we obtain the asymptotic behavior
$\mathcal C/N \simeq k_F^2 \pi/(2 \ln^2(\ad/\lambda))$, which
remarkably resembles the $T=0$ weakly interacting limit where
$\lambda$ is replaced by $1/k_F$.

\begin{figure*}
\centering
\includegraphics[width=\linewidth]{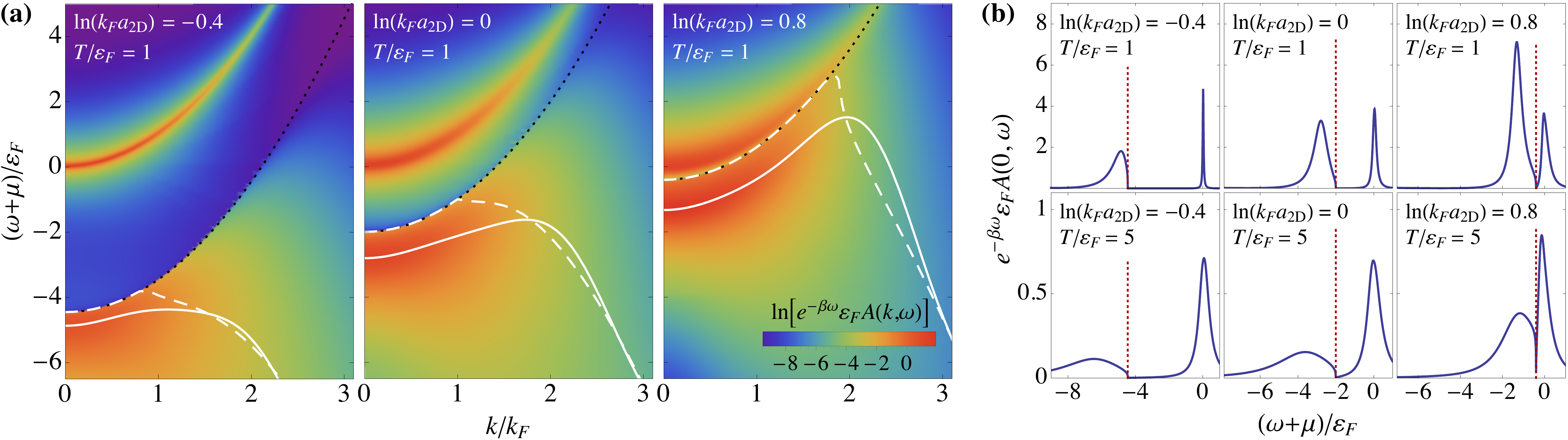}
\caption{(a) The occupied part of the 
spectral function for various values of the interaction
  parameter $\ln (k_F \ad)$ near the crossover. 
They all correspond to expansion parameter $ze^{\beta\eb/2} \simeq 0.7$. 
The (black) dotted line corresponds 
to the free particle dispersion shifted by the binding energy, i.e., $\epsilon_\vect{k}-\eb$, while the (white) 
solid line depicts the position of the peak in the incoherent part of the spectrum at negative energies. The (white) dashed line is the peak position predicted for a thermal gas of bosonic dimers.
(b) Slices of the spectra at $\vect{k}=0$ for two different temperatures. The vertical dotted lines indicate 
the two-body binding energy $-\eb$. 
\label{fig:spectral}}
\end{figure*}

Next, we turn to the spectral function
$A_\sigma(\k,\omega)=-2\text{Im}G_\sigma(\k,\omega)$, which is 
related to the probability of extracting 
an atom in state $\sigma$ with momentum $\k$
and frequency $\omega$. Thus, it is of great relevance to current
experiments measuring the spectral response with momentum resolved photoemission spectroscopy~\cite{Stewart2008,Feld2011}. 
The spectral function is obtained from the full Green's function;
this is related to the bare propagator via the Dyson
equation 
$G^{-1} = [G^{(0)}]^{-1} - \Sigma$,
where $\Sigma$ is the self-energy.
The lowest non-trivial order of the virial expansion yields 
\begin{align} \notag
\Sigma_\up(\p,\tau)=
\begin{tikzpicture}[baseline=(current  bounding  box.center)]
    \filldraw[fill=gray!20]    (0,0)    rectangle    (0.5,0.5);
    \node at             (0.25,0.25) {$T_2$};
    \draw[fermi, decoration={markings,mark=at position .6 with {\arrow[line width=0.3pt]{|}}}]  
                    (.5,.5)    arc    (-40:220: .33 and .25);
    \node at             (0.25,-0.25) {};
\end{tikzpicture}
=ze^{\mu\tau}\sum_\k T_2(\k,\tau)G_\down^{(0,1)}(\k-\p,-\tau),
\end{align}
which clearly includes two-body correlations. Considering the {\em occupied} part of the spectral function introduces a further power of $z$ from the distribution function, and thus the self energy at this order is equivalent to the second order of the virial expansion.
As well as being accurate for $T > \ef$, our results for the contact suggest that the truncation to second order is also reasonable for $T \simeq \ef$ when the interaction is not too weak. Indeed, this lowest-order spectral function has been applied to even lower temperatures in 3D~\cite{Hu2010}.

Figure~\ref{fig:spectral} displays a series of spectra for $T/\ef
\gtrsim 1$ and for a range of $\ln(k_F\ad)$ 
similar to that used in the recent 2D photoemission
experiments~\cite{Feld2011}.  In Fig.~\ref{fig:spectral}(a), we see a
two branch structure that resembles that observed in
experiment~\cite{Feld2011}. The lowest incoherent band corresponds to
bound dimers while the upper branch corresponds to excited unpaired
atoms. A key point is that the spectra all have $\mu < 0$ and so the
effects of Pauli blocking are minimal.  While the temperatures here
are higher than those in experiment, we note that all interactions
except $\ln(k_F\ad) = 0.8$, the lowest interaction considered in
experiment, have $\mu < 0$ at $T=0$. Furthermore, $\ln(k_F\ad) = 0.8$
only has $\mu \simeq 0.2 \ef$ at $T=0$ and thus the Fermi gas is no
longer degenerate
once $T \gtrsim 0.2\ef$, i.e., for typical experimental temperatures.
{Note, further, that our statements appear to be unaffected by the presence of a harmonic trapping potential, since we find that the average chemical potential at $T=0$ in the trap also vanishes at $\ln(k_F\ad)\simeq 0.5$, where $k_F$ is defined as in Ref.~\cite{Frohlich2011}.}

Therefore, our results suggest that the observed pairing
gap~\cite{Feld2011} effectively arises from two-body physics and does
not correspond to a pseudogap regime.  This view is further supported
by the fact that the pairing gap in the spectrum persists to very high
temperatures well above $T_c$, as shown in Fig.~\ref{fig:spectral}(b).
Moreover, we see that the ``closure'' of the gap with increasing
temperature appears to be due to the thermal broadening of the two
branches.

By examining the peak position of the incoherent branch, we obtain a
dispersion that exhibits a ``back-bending'' feature, as shown in
Fig.~\ref{fig:spectral}(a). A similar feature has been observed in 3D
and has been interpreted as evidence for a gapped Fermi
surface~\cite{Gaebler2010}.  However, in our case, it appears to be a
consequence of short-range pairing correlations, similar to what was
described in Ref.~\cite{Schneider2010}. Moreover, the momentum at
which the back-bending occurs is set by $\ad$ and $\lambda$ rather
than $k_F$.

We can compare our results with the atom spectrum obtained for 
a thermal gas of dimers. Here, one probes the spectrum by exciting one of the atoms within the
dimer into a non-interacting state, effectively dissociating the
dimer. The resulting (occupied part of the) spectral function is then proportional to
\begin{align}\notag
& \frac{\mathcal C_0}{m^2} \Theta\left(\frac{k^2}{2m} - \varepsilon_B - (\omega+\mu)\right) \frac{e^{-\beta(3\frac{k^2}{2m}-\varepsilon_B - (\omega+\mu))}}{(\frac{k^2}{2m}- (\omega+\mu))^2} 
\\ 
& \times I_0\left(2\beta \frac{k}{\sqrt{m}}\sqrt{\frac{k^2}{2m}-\varepsilon_B - (\omega + \mu)}\right)
\end{align}
where $I_0(x)$ is a modified Bessel function of the first kind, and $\mathcal C_0 = 2\pi N/\ad^2$ is the two-body contact. This yields a back-bending feature similar to the one obtained in the virial expansion, as shown in Fig.~\ref{fig:spectral}(a), and we 
clearly see how the short-range correlations appear in $A(\vect{k},\omega)$.
Note that the incoherence of the lower band is due to the fact that we have a thermal gas of dimers. If they had been condensed, then we would expect a strongly peaked signal at $\omega + \mu = -\eb -\epsilon_\vect{k}$.

Finally, we note that the virial coefficients obtained above for the
uniform system may be straightforwardly related to those for an
isotropic harmonic confinement~\cite{Liu2010} by means of a local
density approximation: in the thermodynamic limit where the
temperature is much larger than the trapping frequency $\omega$, we
define the local chemical potential
$\mu(\rho)\equiv\mu-\frac12m\omega^2\rho^2$, with $\rho$ the radial
coordinate in the trap. Then the fugacity also depends on the distance
from the center of the trap, $z(\rho) \equiv z
\exp(-\frac12m\omega^2\rho^2\beta)$. Importantly, the virial
coefficients are constant over the trap, and the total number of
particles in the trap is easily obtained by integrating
Eq.~(\ref{eq:dens}) to yield
$N=\frac{2}{\beta^2\omega^2}\sum_{j\geq1} b_j z^j$. On the other
hand, Ref.~\cite{Liu2010} calculates the thermodynamic potential in
the isotropic harmonic trap,
$\Omega=-\frac{2}{\beta^3\omega^2}\sum_{j\geq1} b_j^\text{trap}
z^j$. Computing the particle number $N=-\left.\del \Omega/\del
  \mu\right|_{T,V}$, we identify $b_j=jb_j^\text{trap}$. Using this relation,
we conclude that our virial coefficients match exactly those
of Ref.~\cite{Liu2010}.

In conclusion, we have investigated the temperature dependence of the
contact and the spectral function for the normal 2D Fermi gas.  We
have argued that current experiments are mainly in the Bose regime and
are thus unlikely to have observed pairing that is strongly modified by
Fermi statistics. It remains to be seen whether pseudogap phenomena
could be observed for weaker interactions in 2D~\cite{Bauer2013}.
An unequivocal signature would be the presence of a 
  gap-like feature in the dispersion at finite momentum, occurring at the chemical potential.
In the future, it would be interesting to consider the mass imbalanced
case, where higher partial waves are important in the three-body
problem~\cite{Ngamp2013} and the third virial coefficient can become
non-monotonic~\cite{Daily2012}.

\acknowledgments
We thank M.~Bauer, N.~R.~Cooper, and X.~Leyronas for stimulating
discussions. X.-J.~Liu, H.~Hu, and P.~D.~Drummond are thanked for
sharing their data on the 2D virial coefficients in a trap.
JL acknowledges support from the Carlsberg Foundation
and MMP acknowledges support from the EPSRC under Grant No.\ EP/H00369X/2. 
This work was supported in part by the National Science Foundation under Grant No.~PHYS-1066293 and the hospitality of the Aspen Center for Physics.

\bibliography{virial}

\end{document}